\documentclass[aoas,MSNbibl,nameyear,rotating,dvips]{arximspdf}
\usepackage{graphicx}

\doi{10.1214/11-AOAS532}
\volume{6}
\issue{3}
\pubyear{2012}
\firstpage{1209}
\lastpage{1235}

\makeatletter
\newcommand{\eqref}[1]{(\ref{#1})}
\makeatother

\begin{document}
\begin{frontmatter}

\title{Network inference and biological dynamics\thanksref{T1}}
\runtitle{Network inference and biological dynamics}

\thankstext{T1}{Supported in part by EPSRC EP/E501311/1 (CJO \& SM)
and NCI U54 CA 112970 (SM) and the Cancer Systems Biology Center grant from the
Netherlands Organisation for Scientific Research.}

\begin{aug}
\author[A]{\fnms{Chris. J.} \snm{Oates}\corref{}\ead[label=e1]{c.j.oates@warwick.ac.uk}\ead[label=e3]{c.oates@nki.nl}}
\and
\author[B]{\fnms{Sach} \snm{Mukherjee}\ead[label=e2]{s.n.mukherjee@warwick.ac.uk}\ead[label=e4]{s.mukherjee@nki.nl}}
\runauthor{C. J. Oates and S. Mukherjee}
\affiliation{University of Warwick and Netherlands Cancer Institute, and~Netherlands~Cancer Institute and University of Warwick}

\address[A]{Centre for Complexity Science \\
University of Warwick \\
Coventry, CV4 7AL\\
United Kingdom\\
\printead{e1}\\
and\\
Department of Biochemistry\\
Netherlands Cancer Institute\\
1066CX Amsterdam\\
Netherlands\\
\printead{e3}} 
\address[B]{Department of Statistics\\
University of Warwick\\
Coventry, CV4 7AL\\
United Kingdom\\
\printead{e2}\\
and\\
Department of Biochemistry\\
Netherlands Cancer Institute\\
1066CX Amsterdam\\
Netherlands\\
\printead{e4}}
\end{aug}

\received{\smonth{6} \syear{2011}}
\revised{\smonth{11} \syear{2011}}

%
\begin{abstract}

Network inference approaches are now widely used in
biological applications to probe regulatory
relationships between molecular components such as genes or proteins.
Many methods have been proposed for this setting, but the connections
and differences between their statistical formulations have received less
attention.
In this paper, we show how a broad class of statistical
network inference methods, including a number of existing approaches,
can be described in terms of variable selection for the linear model.
This reveals
some subtle but important differences between the methods, including
the treatment of time intervals in discretely observed data.
In developing a~general formulation, we also explore the relationship
between single-cell stochastic dynamics and network inference on
averages over cells. This clarifies the
link between biochemical networks as they
operate at the cellular level and network inference as
carried out on data that are averages over populations of cells.
We present empirical results, comparing thirty-two network
inference methods that are instances of the general formulation we
describe, using two published dynamical models.
Our investigation sheds light on the applicability and limitations of
network inference and provides guidance for practitioners and
suggestions for experimental design.
\end{abstract}

%
\begin{keyword}
\kwd{Network inference}
\kwd{biological dynamics}
\kwd{variable selection}
\end{keyword}

\end{frontmatter}
%

\section{Introduction}

Networks of molecular components such as genes, proteins and
metabolites play a prominent role in molecular biology.
A graph $G = (V,E)$ can be used to describe a biological network, with
the vertices $V$ identified with molecular components
and the edges $E$ with regulatory relationships between them.
For example, in a gene regulatory network [\citet{Babu}; \citet{Davidson}],
nodes represent genes and edges transcriptional regulation, while in a
protein signaling network [\citet{Yarden}], nodes represent proteins and
edges may represent the enzymatic influence of the parent on the
biochemical state of the child, for example, via phosphorylation.
In many biological contexts, including disease states, the edge
structure of the network may itself be uncertain (e.g., due to genetic
or epigenetic alterations). Then, an important biological goal is to
characterize the edge structure (often referred to as the ``topology''
of the network) in a context-specific manner, that is, using data
acquired in the biological context of interest (e.g., a type of cancer,
or a developmental state).
Advances in high-throughput data acquisition have led to much interest
in such data-driven characterization of biological networks.
Statistical approaches play an increasingly important role in these
``network inference'' efforts.
From a statistical perspective, the goal can be viewed as making
inference regarding the edge structure $E$ in light of
biochemical data $\mathbf{y}$.
Since aspects of biological
dynamics may not be identifiable at steady-state, time-varying data is
usually preferred, and this is the setting we focus on here.
In many applications the data $\mathbf{y}$ arise from ``global
perturbation'' of the cellular system, for example, by varying culture
conditions or stimuli. The extent to which networks can be
characterized using global perturbations remains poorly understood,
since it is likely that such data expose only a~subspace of the phase
space associated with cellular dynamics.

The importance of network inference in diverse biological
applications, from basic biology to diseases such as cancer, has
spurred vigorous activity in this area. Many specific methods have
been proposed, in the statistical literature as well as in
bioinformatics and bioengineering, with some popular approaches
reviewed in
\citet{BansalII}; \citet{Bonneau}; \citet{Hecker}; \citet{Lee}; \citet{Markowetz}.
Graphical models play a prominent role in this literature, as does
variable selection.
A distinction is often made between statistical and ``mechanistic''
approaches [\citet{Ideker}].
The former is usually used to refer to models that are built on
conventional regression formulations and variants thereof, while the
latter usually refers to models that are explicitly rooted in chemical
kinetics, for example, systems of coupled ordinary differential equations
(ODEs).
This distinction is somewhat artificial, since it is possible in
principle to carry out formal statistical network inference based on
mechanistic models (e.g., systems of ODEs), although this remains
challenging [\citet{Xu}].

Many network inference schemes are based on formulations that are
closely related in terms of the underlying statistical model. For
example, vector autoregressive (VAR) models [including Granger
causality-related approaches as special cases; \citet
{Bolstad}; \citet{Meinshausen}; \citet{Morrissey}; \citet{Opgen}; \citet{Zou}],
linear dynamic Bayesian networks [DBNs; \citet{Kim}] and
certain\vadjust{\goodbreak}
ODE-based approaches
[Bansal and di~Ber\-nardo (\citeyear{Bansal}); \citet{Li}; \citet{Nam}] are intimately related, being based
on linear regression, but with potentially differing approaches to
variable selection.
In recent years, several empirical comparisons of competing network
inference schemes have emerged, including
\citet{Altay}; \citet{BansalII}; \citet{Hache}; \citet{Smith}; \citet{Werhli}.
Assessment methodology has received attention, including attempts to
automate the generation of large scale biological network models for
automatic benchmarking of performance [\citet{Marbach}; \citet{Bulcke}]. In
particular, the Dialogue for Reverse Engineering Assessments and
Methods (DREAM) challenges [\citet{Prill}] have provided an opportunity
for objective empirical assessment of competing approaches.
At the same time, developments in synthetic biology have led to the
availability of gold standard data from hand-crafted biological
systems, such that the underlying network is known by design
[\citet{Camacho}; \citet{Cantone}; \citet{Minty}].
However, relatively little attention has been paid to the (sometimes
contrasting) assumptions of the statistical formulations underlying
these network inference schemes.

Inferential limitations due to estimator bias and nonidentifiability
remain incompletely understood. It is clear that chemical reaction
networks (CRNs; these are graphs that give detailed descriptions of
individual reactions comprising the overall system) underlying
biological networks are not in general identifiable [\citet{Craciun}].
Indeed, there exist topologically
distinct CRNs which produce identical
dynamics under mass-action kinetics. Moreover, even when the true
network structure is known, reaction rates
themselves may be nonidentifiable.
However, mainstream descriptions of biological networks, for example,
gene regulatory or protein signaling networks, are coarser than CRNs.
Such networks are useful because they are closely tied to validation
experiments in which interventions (e.g., RNA interference or
inhibitors) target network vertices. For example, inference of an edge
in a~gene regulatory network corresponds to the qualitative prediction
that intervention on the parent will influence the child (via
transcription factor activity).
It remains unclear to what extent such biological network structure can
be usefully identified from various kinds of data.
On the other hand, \citet{Wilkinson}; \citet{WilkinsonII} discusses a number
of general issues relating to stochastic modeling for systems
biology, but does not discuss network inference \textit{per se} in detail.
This paper complements existing empirical work by focusing on
statistical issues associated with linear models commonly used in network
inference applications.

Network inference methods can be viewed
as generating hypotheses about cell biology. Yet the link between
biochemical networks at the cellular level and network inference as
applied to bulk or aggregate data (i.e., data that are averages
over\vadjust{\goodbreak}
large numbers of cells) from assays such as microarrays remains
unclear.
In applications to noisy time-varying data there is uncertainty in the
predictor variables of the same order of magnitude as uncertainty in
the responses, yet often only the latter is explicitly accounted for.
Moreover, the treatment of time intervals in discretely observed data
remains unclear, with contradictory approaches appearing in the
literature.
Most high-throughput assays, including array based technologies (e.g.,
gene expression or protein arrays), as well as single-cell approaches
(e.g., FACS-based) involve destructive sampling, that is, cells are
destroyed to obtain the molecular measurements.
The impact of the resulting nonlongitudinality upon inference
does not appear to have been investigated.

The contributions of this paper are threefold. First, we explore the
connection between biological networks at the cellular level and the
linear statistical models that are widely used for inference.
Starting from a description of stochastic dynamics at the single-cell
level, we describe a general statistical approach rooted in the linear
model.
This makes explicit the assumptions that underlie a broad class of
network inference approaches.
This also clarifies the relationship between ``statistical'' and
``mechanistic'' approaches to biological networks.
Second, we explore how a number of published network
inference approaches can be recovered as special
cases of the model we arrive at. This sheds light on the
differences between them, including how different assumptions lead to
quite different treatments of the time step.
Third, we present an empirical study comparing 32 different approaches
that are special cases of the general model we describe. To do so, we
simulate stochastic dynamics at the single-cell level from known
networks, under global perturbation of two published dynamical models.
This enables a clear assessment of the network inference methods in
terms of estimation bias and consistency,
since the true data-generating network is known. Furthermore, the
simulation accounts for both averaging over cells, nonlongitudinality
due to destructive sampling and the fact that only a subspace of the
dynamical phase space is explored.
Using this approach, we investigate a number of data regimes, including
both even and uneven sampling, longitudinal and nonlongitudinal data
and the large sample, low noise limit.
We find that the net effect of
predictor uncertainty, nonlongitudinality and limited exploration of
the dynamical phase space is such that
certain network estimators fail to converge to the data-generating
network even in the limits of
large data sets and low noise.
However, we point to a simple formulation which might
represent a default choice, delivering promising performance in a
number of regimes.

An implication of our analysis is that uneven time steps may pose
inferential problems, even when using models that apparently handle
the sampling intervals explicitly. We therefore investigate this case
by carrying out network inference on unevenly sampled data using a
variety of statistical models. We find that the ability to reconstruct
the data-generating network is much reduced in all cases, with some
approaches faring\vadjust{\goodbreak} better than others. Since biological data are often
unevenly resolved in time, this observation has important implications
for experimental design.

The remainder of this paper is organized as follows. We begin in
Section~\ref{sec: methods} with a description of stochastic dynamics
in single cells and show how a series of assumptions allow us to
arrive at a statistical framework rooted in the linear model.
Section~\ref{sec: Results} contains an empirical comparison of several
inference schemes, addressing questions of performance and consistency
in a number of regimes.
In Section~\ref{sec: discuss} we
discuss our results and point to several specific areas for future work.

\section{Methods} \label{sec: methods}

The cellular dynamics that underlie network inference are subject to
stochastic effects
[\citet{Elowitz}; \citet{Kou}; \citet{McAdams}; \citet{Paulsson}; \citet{Swain}].
We therefore begin our description of the data-generating process at
the level of single cells and then discuss the relationship to
aggregate data of the kind acquired in high-throughput biochemical
assays. We then develop a general statistical approach, rooted in the
linear model, for data from such a system observed discretely in time.
We discuss inference and show how
a number of existing approaches can be recovered as special
cases of the general model we describe. Our exposition clarifies a
number of technical but important distinctions between published
methodologies, which until now have received little attention.

\subsection{Data-generating process}
\subsubsection{Stochastic dynamics in single cells}
Let $\mathbf{X} = (X_1,\ldots,X_P) \in\mathcal{X}$ denote a state
vector describing
the abundance of molecular quantities of interest, on a space $\mathcal
{X}$ chosen according to physical and statistical considerations.
The components of the state vector (e.g., mRNA, protein or metabolite
levels) are identified with the vertices of
the graph $G$ that describes the biological network of interest.
In this paper the ``expression levels'' $\mathbf{X}(t)$ of a single
cell at time $t$
are modeled as continuous random variables that we
assume satisfy a time-homogenous stochastic delay differential
equation (SDDE)
%
\begin{equation}
d\mathbf{X}  = \mathbf{f}(\mathcal{F}_{\mathbf{X}})\,dt +
\mathbf{g}(\mathcal{F}_{\mathbf{X}})\,d\mathbf{B},
\label{eq: single cell}
\end{equation}
where $\mathbf{f},\mathbf{g}$ are drift and diffusion functions
respectively,
\mbox{$\mathcal{F}_{\mathbf{X}}(t) = \{\mathbf{X}(s) : s \leq t \}$}
is the natural filtration (the history of the state vector
$\mathbf{X}$) and $\mathbf{B}$ denotes a standard Brownian motion. A
continuous state space $\mathcal{X}$ is appropriate as a~modeling
assumption only if the copy numbers of all molecular components are
sufficiently high. This is thought to be the case for the biological
systems considered in this paper, but in general the stochasticity due
to low copy number will need to be encoded into inference [\citet
{Paulsson}]. The edge structure $E$ of the biological network $G$ is
defined by the drift function~$\mathbf{f}$, such that $(i,j)\in E \iff
f_j(\mathbf{X})$ depends on $X_i$.\vadjust{\goodbreak}

We further assume that the functions $\mathbf{f},\mathbf{g}$ are sufficiently
regular and depend only on recent history $\mathcal{F}_{\mathbf{X}}
([t-\tau,t])$. For example, in the context of gene regulation $\tau$
might be the time required for one cycle of transcription, translation
and binding of a transcription factor to its target site, the
characteristic time scale for gene regulation.
This is a finite memory requirement and can be considered a
generalization of the Markov property. Equivalently, this property
codifies the modeling assumption that the observed processes are
sufficient to explain their own dynamics, that there are no latent variables.
It is common practice to take $\tau= 0$, in which case the process
defined by equation (\ref{eq: single cell}) is Markovian. This stochastic
dynamical system with phase space $\{ (\mathbf{f}(\mathcal{F}_{\mathbf
{X}}),\mathbf{X}) : \mathbf{X}\in\mathcal{X} \}$ forms the basis of the
following exposition.

\subsubsection{Aggregate data}
A variety of experimental techniques, including, notably, microarrays
and related assays,
capture average expression levels $\mathbf{X}^{(N)}
:= \sum_{k=1}^N\mathbf{X}^k / N$ over cells, where $\mathbf{X}^k$ denotes
the expression levels in cell $k$. This paper does not consider effects
due to intercellular signaling, which are typically assumed to be
negligible. Then averaging sacrifices the finite
memory property (a~generalization of the fact that the sum of two
independent Markov processes is not itself Markovian). However, it is usually
possible to construct a finite memory approximation of the form
%
\begin{equation}
d\mathbf{X}^{(N)}  = \mathbf{f}^{(N)}\bigl(\mathcal{F}_{\mathbf
{X}^{(N)}}\bigr)\,dt +
\mathbf{g}^{(N)}\bigl(\mathcal{F}_{\mathbf{X}^{(N)}}\bigr)\,d\mathbf{B}^{(N)}
\label{eq: SDE model}
\end{equation}
using a so-called ``system size expansion'' [\citet{Kampen}].
Approximations of this kind derive from a coarsening of the underlying
state space, assuming that the new state vector $\mathbf{X}^{(N)}$
captures every quantity relevant to the dynamics. The statistical
models discussed in this paper rely upon coarsening assumptions in
order to control the dimensionality of state space.

Using the mild regularity conditions upon cellular stochasticity
$\mathbf{g}$, the laws of large numbers give that in
the large sample limit the sample average $\mathbf{X}^\infty:= \lim
_{N\rightarrow\infty}
\mathbf{X}^{(N)} = \mathbb{E}(\mathbf{X})$ equals the expected state of
a single cell (almost surely). We note that the relationship
between the single-cell dynamics as it appears in equation (\ref{eq: single
cell}) and this deterministic limit may be complicated, since in general
$\mathbb{E}(\mathbf{f}(\mathcal{F}_{\mathbf{X}})) \neq\mathbf
{f}(\mathcal{F}_{\mathbb{E}
(\mathbf{X})})$. However, for linear
$\mathbf{f}$, say, for simplicity, $\mathbf{f} \equiv\mathbf{f}(\mathbf
{X}) = \mathbf{AX}$, we have
%
\begin{eqnarray}
d\mathbf{X}^{(N)} &=& \frac{1}{N}\sum_{k=1}^N d\mathbf{X}^k
 =  \frac{1}{N}\sum_{k=1}^N \bigl( \mathbf{f}(\mathcal{F}_{\mathbf
{X}^k})\,dt + \mathbf{g}(\mathcal{F}_{\mathbf{X}^k})\,d\mathbf{B}^k \bigr)
 \nonumber\\
& = & \frac{1}{N}\sum_{k=1}^N \mathbf{AX}^kdt + \frac{1}{N}\sum_{k=1}^N
\mathbf{g}(\mathcal{F}_{\mathbf{X}^k})\,d\mathbf{B}^k
\nonumber
\\[-8pt]
\\[-8pt]
\nonumber
& = & \mathbf{A}\Biggl(\frac{1}{N}\sum_{k=1}^N \mathbf{X}^k\Biggr)\,dt +
\mathbf{R}^{(N)} \\
& = & \mathbf{A}\mathbf{X}^{(N)}\,dt + \mathbf{R}^{(N)} = \mathbf
{f}\bigl(\mathcal{F}_{\mathbf{X}^{(N)}}\bigr)\,dt + \mathbf{R}^{(N)}, \nonumber
\end{eqnarray}
where $\mathbf{R}^{(N)}:= \sum_k \mathbf{g}(\mathcal{F}_{\mathbf
{X}^k})\,d\mathbf{B}^k /N \rightarrow\mathbf{0}$ almost surely as
$N\rightarrow\infty$, and so $d\mathbf{X}^\infty/dt = \mathbf{f}
(\mathcal{F}_{\mathbf{X}^\infty})$. In other words, the average over
large numbers of cells shares the same drift function as the single
cell, so that inference based on averaged data applies directly to
single-cell dynamics. Otherwise this may not
hold, that is, $d\mathbf{X}^\infty/dt = d\mathbb{E}(\mathbf{X})/dt =
\mathbb{E}(\mathbf{f}(\mathcal{F}_{\mathbf{X}})) \neq\mathbf
{f}(\mathcal{F}_{\mathbb{E}(\mathbf{X})}) =
\mathbf{f}(\mathcal{F}_{\mathbf{X}^\infty})$. This has implications
when using nonlinear forms, such as Michaelis-Menten or Hill kinetics,
to describe
the behavior of a large sample average; these nonlinear functions are
derived from single-cell biochemistry and may not apply equally to the
large sample average $\mathbf{X}^\infty$. The error entailed by
commuting drift and expectation may be assessed using the multivariate
Feynman-Kac formula for $\mathbf{X}^{\infty} = \mathbb{E}(\mathbf{X})$
[\citet{Oksendal}].

In practice, the observation process may be complex and indirect, for
example, measurements of gene expression may be relative to a
``housekeeping'' gene, assumed to maintain constant expression over the
course of the experiment. Moreover, the details of the error structure
will depend crucially on the technology used to obtain the data. To
limit scope, this article assumes the averaged expression levels
$\mathbf{X}^{\infty}(t)$ are observed at discrete times $t = t_j$
($0\leq
j\leq n$) with additive zero-mean measurement error as
$\mathbf{Y}(t_j) = \mathbf{X}^{\infty}(t_j) + \mathbf{w}_j$, where the
$\mathbf{w}_j$ are independent, identically distributed uncorrelated
Gaussian random variables.

\subsection{Discrete time models} \label{sec: assumptions}

Network inference is usually carried out using coarse-grained models
[equation (\ref{eq: SDE model})] that are simpler and more amenable to
inference than the process
described by equation (\ref{eq: single cell}).
Here, informed by the foregoing treatment of cellular dynamics, we
develop a simple network inference model for data observed discretely
in time.
We clarify the assumptions of the statistical model, and show how
several published approaches can be recovered as special cases.

\subsubsection{Approximate discrete time likelihood}
Network inference entails statistical comparison of networks $G \in
\mathcal{G}$, where $\mathcal{G}$ denotes the space of candidate
networks.
The space $\mathcal{G}$ may be large (naively, there are $2^{P\times
P}$ possible networks on $P$ vertices), although biological knowledge
may provide
constraints.
Network comparisons require computation of a model selection score
for each network, that is, considered, which in turn entails use of the
likelihood (e.g., maximization of information criteria, or integration
over the likelihood in the Bayesian
setting). Therefore,
exploration over large model spaces is often only feasible given a
closed-form expression for the likelihood (or preferably for the model
score itself).\vadjust{\goodbreak}

However, the likelihood for a SDDE model [equation (\ref{eq: SDE
model})]
is not generally available in closed form.
There has been recent research into computationally efficient
approximate likelihoods for fully observed, noiseless diffusions [\citet
{Hurn}], but
it remains the case that the most efficient (though least
accurate) closed-form approximate likelihood is based on the
Euler-Maruyama discretization scheme for stochastic differential equations
(SDEs), which in the more general SDDE case may be written as
(henceforth dropping the superscript~$N$)
%
\begin{equation}
\mathbf{X}(t_j)  \approx \mathbf{X}(t_{j-1}) + \Delta_j \mathbf{f}
(\mathcal{F}_{\mathbf{X}}(t_{j-1})) + \mathbf{g}
(\mathcal{F}_{\mathbf{X}}(t_{j-1}))\Delta\mathbf{B}_j,
\label{eq: Euler}
\end{equation}
where $\Delta\mathbf{B}_j \sim N(\mathbf{0},\Delta_j \mathbf{I})$ and
$\Delta_j = t_j - t_{j-1}$ is the sampling time interval.
Incorporating measurement error into this so-called Riemann-It\^{o}
likelihood [\citet{Dargatz}] requires an integral over the hidden states
$\mathbf{X}$
which would destroy the closed-form approximation. Therefore, the
observed, nonlongitudinal data $\mathbf{y}$ are directly substituted
for the latent states
$\mathbf{X}$, yielding the (triply) approximate likelihood
%
\begin{eqnarray} \label{eq: AL 1}
\mathcal{L}(\theta) & = & \prod_{j=1}^n \mathcal{N} (\mathbf{y}
(t_j);\mu(t_j),\Sigma(t_j)),\nonumber\\
\mu(t_j) & = & \mathbf{y}(t_{j-1})+\Delta_j \mathbf{f}
(\mathcal{F}_{\mathbf{y}}(t_{j-1})),  \\
\Sigma(t_j) & = & \Delta_j \mathbf{g}(\mathcal{F}_{\mathbf{y}}(t_{j-
1}))\mathbf{g}(\mathcal{F}_{\mathbf{y}}(t_{j-1}))'.\nonumber
\end{eqnarray}
Here $\mathcal{N}(\bullet;\mu,\Sigma)$ denotes a Normal density with
mean $\mu$ and covariance $\Sigma$. Implicit here is that the
functions $\mathbf{f},\mathbf{g}$ depend on $\mathcal{F}_{\mathbf{y}}$
only through time lags which coincide with the measurement times
$t_{j-1}$.

Thus, $\mathcal{L}$ may be obtained from a state-space approximation to
the original SDDE model [equation (\ref{eq: SDE model})].
Despite reported weaknesses with the Riemann-It\^{o} likelihood
[\citet{Dargatz}; \citet{Hurn}] and the poorly characterized error incurred by
plugging in nonlongitudinal
observations, this form of approximate likelihood is widely used to
facilitate network inference [equations~(\ref{eq: AL 1}) and (\ref{eq: AL 3})
correspond to a Gaussian DBN for the observations $\mathbf{y}$,
generalized to allow dependence on
history]. This is due both to the possibility of parameter
orthogonality, allowing inference to be performed for each network
node separately, and the possibility of conjugacy, leading to a
closed-form marginal likelihood $\pi(\mathbf{y}|G)$.

\subsubsection{Linear dynamics}

Kinetic models have been described for many cellular processes
[\citet{Cantone}; \citet{Schoeberl}; \citet{Swat}; \citet{WilkinsonII}].
However, statistical inference for these often nonlinear models may
be challenging [\citet{Bonneau}; \citet{Wilkinson}; \citet{WilkinsonII}; \citet{Xu}].
Moreover, there is no guarantee that conclusions drawn from cellular
averages will apply to single cells, because,\vadjust{\goodbreak} as noted above, the
deterministic behavior seen in averages may not coincide with the
single-cell drift. However, linear dynamics satisfy $\mathbb{E}(\mathbf{f}
(\mathcal{F}_{\mathbf{X}})) = \mathbf{f}(\mathcal{F}_{\mathbb{E}(\mathbf
{X})})$ exactly, so that conclusions drawn from verages apply directly
to single cells.
For notational simplicity consider the Markovian $\tau=0$ regime. A
Taylor approximation of the cellular drift~$\mathbf{f}$ about the
origin gives
%
\begin{equation}\label{eq: AL 3}
\mathbf{f}(\mathbf{X}) \approx \mathbf{f}(\mathbf{0}) +
D\mathbf{f}|_{\mathbf{x}=\mathbf{0}} \mathbf{X},
\end{equation}
where $D\mathbf{f}$ is the Jacobian matrix of $\mathbf{f}$. The
constant term can be omitted ($\mathbf{f}(\mathbf{0})=\mathbf{0}$),
since absent any regulators there is no change in expression.
Then, the Jacobian $D\mathbf{f}$ captures the dynamics approximately
under a linear model.
Furthermore, the absence of an edge in the network $G$ implies a zero
entry in the Jacobian, that is, $(i,j)\notin E \Rightarrow(D\mathbf
{f})_{ji}=0$. Obtaining the Jacobean at $\mathbf{x}=\mathbf{0}$
therefore does not imply complete knowledge of the edge structure $E$.
We note that the general SDDE case is similar but with additional
differentiation required for the additional dependencies of
$\mathbf{f}$. Henceforth, we write equations for the simpler Markovian
model, although they hold more generally.

One may ask whether the restriction to linear drift functions allows
the computational difficulties associated with inference for
continuous time models to be avoided, since in the Markovian ($\tau=
0$) case both the SDE [equation (\ref{eq: single cell})] and limiting
ordinary differential equation (ODE) have exact closed form solutions.
In the ODE case, for example, $\mathbf{X}(t) =
\exp(\mathbf{A}t)\mathbf{X}_0$ and under Gaussian measurement error
the likelihood has a closed form as products of terms $\mathcal{N}
(\mathbf{y}(t_j);\exp(\mathbf{A}t_j)\mathbf{X}_0,\mathbf{M})$, where
the parameters $\theta= (\mathbf{A},\mathbf{X}_0,\mathbf{M})$
include the model parameters $\mathbf{A}$, initial state vector
$\mathbf{X}_0$ and the measurement error covariance $\mathbf{M}$.
Unfortunately, evaluation of the matrix exponential is computationally demanding
and inference for the entries of $\mathbf{A}$ must be
performed jointly since, in general, $ \exp(\mathbf{A})$ does not factorize
usefully. It therefore remains the case that inference for continuous
time models is computationally burdensome, even when the models are
linear.

\subsubsection{The dynamical system as a regression model}
The Jacobian $D\mathbf{f}$ with entries $(D\mathbf{f})_{i,j} =
\partial
f_i / \partial x_j |_{\mathbf{x} = \mathbf{0}}$ is now the focus
of inference. We can identify the Jacobian with the unknown parameters
in a linear regression problem by modeling the expression of gene $p$
using
%
\begin{equation}
\left[
\matrix{ dX_p(t_1) \vspace*{2pt}\cr \vdots\vspace*{2pt}\cr dX_p(t_n)}
\right] \approx
\left[
\matrix{ X_1(t_0) & \cdots& X_P(t_0) \vspace*{2pt}\cr \vdots& &
\vdots\vspace*{2pt}\cr X_1(t_{n-1}) & \cdots& X_P(t_{n-1})}
\right]
\left[
\matrix{(Df)_{p,1} \vspace*{2pt}\cr \vdots\vspace*{2pt}\cr (Df)_{p,P}}
\right],
\label{eq: regression}
\end{equation}
where the gradients $dX_p(t_j)$ are approximated by finite
differences, in this case $(X_p(t_j) - X_p(t_{j-1})) / \Delta_j$. Our
notation for finite differences should not be confused with the
differentials of stochastic calculus. More
generally, for processes\vadjust{\goodbreak} with memory, the matrix may be augmented with
columns corresponding to lagged state vectors and the vector
$(D\mathbf{f})_{p,\bullet}$ augmented with the corresponding
derivatives of the
drift function $\mathbf{f}$ with respect to these lagged states. To
avoid confusion, we write $\mathbf{A}$ for $D\mathbf{f}$ when
discussing parameters, since the drift $\mathbf{f}$ is unknown.
Similarly, design matrices will be denoted by~$\mathbf{B}$ to suppress
the dependence on the random variables $\mathbf{X}$. So equation (\ref{eq:
regression}) may be written compactly as
%
\begin{equation}
d\mathbf{X}_p \approx \mathbf{B} A_{p,\bullet}'.
\label{eq: regression compact}
\end{equation}
Inference for the parameters $A_{p,\bullet}$ may be performed
independently for each variable $p$.
While equation (\ref{eq: regression compact}) is fundamental for inference,
one can equivalently consider the dynamically intuitive expression
%
\begin{equation}
d\mathbf{X}(t_j)  \approx \mathbf{A} B_{j,\bullet}'.
\label{eq: dynamic model}
\end{equation}

An interesting issue arises from the dual interpretation of the
regression model as a dynamical system [equation (\ref{eq: dynamic model})],
because there are natural restrictions on $\mathbf{A}$ to avoid the
solution tending to infinity. For instance, if the sampling interval
$\Delta$ is constant, then we require $\mathbb{R}(\lambda)\leq0$ for
each eigenvalue~$\lambda$ of $\mathbf{A}+\Delta\mathbf{I}$. The
inference schemes which we discuss do not account for this, because
the condition forces a nontrivial coupling between
rows~$A_{p,\bullet}$, jeopardizing parameter orthogonality.

Finally, the generative model is specified by substituting noisy,
nonlongitudinal
observables $\mathbf{Y}$ for latent variables $\mathbf{X}$ into equation
(\ref{eq: dynamic model}) and stating the dependence of the
approximation error on the sampling interval $\Delta_j$. Under
uncorrelated Gaussian measurement error we arrive at a model
%
\begin{equation}
d\mathbf{Y}(t_j)  \sim N(\mathbf{A}
B_{j,\bullet}',h(\Delta_j)\mathcal{D}(\sigma_1^2,\ldots,\sigma_P^2)),
\label{eq: model}
\end{equation}
where $h:\mathbb{R}^+\rightarrow\mathbb{R}^+$ is a variance function
that must be specified and
$\mathcal{D}(\mathbf{v})$ represents the diagonal matrix induced by
the vector $\mathbf{v}$.

There are a number of ways in which this regression is nonstandard.
For example, the substitution of (nonlongitudinal) observations for
latent variables is
clearly unsatisfactory because the linear regression framework does
not explicitly allow for uncertainty in the predictor variables $\mathbf
{B}$. It
is unclear whether this introduces bias or leads to an overestimate of
the significance of results. Moreover, it is unclear how to choose the
variance function $h$, since the Euler-Maruyama approximation [equation
(\ref{eq: Euler})] is only valid for small sampling intervals
$\Delta_j$, but in this regime the responses $d\mathbf{Y}(t_j)$ are
dominated by measurement error, such that the data may carry little
information. These issues are investigated in Sections~\ref{sec:
Results} and~\ref{sec: discuss} below.

\subsection{A unifying framework}
Equation (\ref{eq: model}) describes a class of models with specific
instances characterized by choice of design matrix $\mathbf{B}$ and
variance function~$h$.
Since any such model corresponds to the linear regression equation
(\ref{eq: regression}), the task of determining the edge structure of
the network, or, equivalently, the location of nonzero entries in the
Jacobian $\mathbf{A}$, can be cast as a variable selection problem.\vadjust{\goodbreak}

A number of specific network inference schemes can now be recovered by
fixing the design matrix and variance function and coupling the
resulting model with a variable selection technique.
A selection of published network inference schemes that can be viewed in
this way is presented in Table~\ref{fig: literature}. One might see
these schemes classed as VAR models
[\citet{Bolstad}; \citet{Morrissey}; \citet{Opgen}; \citet{Zou}], DBNs
[\citet{Hill}; \citet{Kim}] or ODE-based
approaches [\citet{Bansal}; \citet{Li}; \citet{Nam}], although as we have
demonstrated this classification disguises their shared foundation in
the linear model.

\begin{table}
\tabcolsep=0pt
\caption{A nonexhaustive list of network inference
schemes rooted in the linear model. The~examples from literature
demonstrate the statistical features indicated, but~may~differ~in~some
aspects of implementation.
The symbol $\varnothing$ denotes
the~$\operatorname{VAR}(q)$~model~which~lacks~a~variance function}\label{fig: literature}
{\fontsize{8.5pt}{10.5pt}\selectfont{\begin{tabular*}{\textwidth}{@{\extracolsep{\fill}}lccc@{}}
\hline
& \textbf{Variance} &  &  \\
\textbf{Design}  & \multicolumn{1}{c}{\textbf{function}}  &  & \\
\multicolumn{1}{@{}l}{\textbf{matrix} $\mathbf{B}$}&\multicolumn{1}{c}{$\mathbf{h}(\bolds{\Delta}) \bolds{\propto}$}&\textbf{Variable selection}&\textbf{Example}\\
\hline
Standard & $\Delta^{-2}$ & Ridge regression & \citet{Bansal} ``TSNIB''
\\
Standard with  & $\varnothing$ & Group LASSO & \citet
{Bolstad} \\
\quad lagged predictors&&&\\
Quadratic & $\varnothing$ & Conjugate Bayesian & \citet{Hill} \\
& & with network prior & \\
Standard & $\varnothing$ & Information criteria & \citet{Kim} \\
Nonlinear (Hill)  & 1 & AIC with backstepping & \citet{Li} \\
\quad basis functions&&&\\
Standard & 1 & Conditional independence  & \citet{LiLi} ``DELDBN''
\\
&&tests&\\
Standard & $\varnothing$ & Semi-conjugate Bayesian & \citet{Morrissey} \\
Standard & $\Delta^{-2}$ & SVD and pseudoinverse & \citet{Nam} ``LEARNe'' \\
Standard & $\varnothing$ & Multi-stage analytic & \citet{Opgen} \\
& & shrinkage approach & \\
Standard and   & $\varnothing$ & Granger causality & \citet{Zou} \\
\quad nonlinear with  & & & \\
\quad lagged predictors&&&\\
\hline
\end{tabular*}}}
\end{table}

As shown in Table~\ref{fig: literature}, the variance functions $h$,
and therefore sampling intervals $\Delta_j$, are not treated in a
consistent way in the literature.
In the special case of even sampling times $\Delta_j = \Delta$, a
model is characterized only by its design matrix. If the standard
design matrix is used, then the entire family of models
%
\begin{equation}
\frac{\mathbf{Y}(t_j)-\mathbf{Y}(t_{j-1})}{\Delta} \sim N(\mathbf
{AY}(t_{j-1}),h(\Delta)\mathcal{D}(\sigma_1^2,\ldots,\sigma_P^2))
\end{equation}
reduces to a
linear $\operatorname{VAR}(1)$ model
%
\begin{equation}
\mathbf{Y}(t_j) \sim N(\bar{\mathbf{A}}\mathbf{Y}(t_{j-1}),\mathcal{D}
(\bar{\sigma}_1^2,\ldots,\bar{\sigma}_P^2)),
\end{equation}
where $\bar{\mathbf{A}} = \Delta\mathbf{A} + \mathbf{I}$ and
$\bar{\sigma}_p^2 = \Delta^2 h(\Delta)\sigma_p^2$.
More generally, the $\operatorname{VAR}(q)$ model is prevalent in the literature (see
Table~\ref{fig: literature}), yet it does not explicitly handle uneven
sampling intervals.
This is a potentially important issue since uneven sampling is
commonplace in global perturbation experiments, with high frequency
sampling used to capture short term cellular response and low frequency
sampling to capture the approach to equilibrium.
We discuss the importance of modeling using a variance function, and
whether a natural choice for such a function exists in Section~\ref{sec: discuss} below.
In addition, we explored whether inference may be improved through the
use of either nonlinear basis functions or lagged
predictors to capture respectively nonlinearity and memory in the
underlying drift function is unclear. Section~\ref{sec: Results}
presents an empirical investigation of these issues.

\subsection{Inference} \label{sec: inference}
An appealing feature of the discrete time model is that parameters
corresponding to different variables are orthogonal in the Fisher sense:
%
\begin{equation}
\mathcal{L}(\theta) = \prod_{p=1}^P \mathcal{L}(A_{p,\bullet},\sigma_p).
\end{equation}
As a consequence,
network inference over $\mathcal{G}$ may be factorized into $P$
independent variable selection problems.
For definiteness we focus on just two approaches to variable selection,
the Bayesian
marginal likelihood and AIC, but note that many other approaches are
available, including those listed in Table~\ref{fig: literature}, and
can be applied here in analogy to what follows.
Below we assume the response vector $d\mathbf{y}_p h^{-1/2}$ and the
columns of the design matrix $\mathbf{B}h^{-1/2}$ are standardized to
have zero mean and unit variance, but for clarity subsume this into
unaltered notation.

\subsubsection{Bayesian variable selection} \label{subsubsec: Bayes}
For simplicity, the variance function is initially taken to be
constant ($h=1$).
We set up a Bayesian linear model conditional on a network $G$ using
Zellner's g-prior [\citet{Zellner}], that is, with priors
$A_{p,\bullet}| \sigma_p^2 \sim N(\mathbf{0},\sigma_p^2 n (
{\mathbf{B}'_p}  \mathbf{B}_p )^{-1})$ and
$\pi(\sigma_p^2) \propto1/\sigma_p^2$
where $\mathbf{B}_p$ is the design matrix $\mathbf{B}$ with
nonpredictors removed according to $G$. We note that while the g-prior is
a common choice, alternatives may offer some advantages
[\citet{Deltell}; \citet{Friedman}].

Let $m_p$ be the number of predictors for variable $p$ in the network
$G$. Integrating the likelihood [induced by equation (\ref{eq: model})]
against the prior for $(A_{p,\bullet},\sigma_p^2)$ produces the
following closed-form marginal likelihood:
%
\begin{equation}
\pi(\mathbf{y}|G)  \propto \prod_p \biggl(\frac{1}
{1+n}\biggr)^{m_p/2} \biggl[ d\mathbf{y}_p'\,d\mathbf{y}_p -
\biggl(\frac{n}{1+n}\biggr) \,\hat{d\mathbf{y}_p}{}^{\prime}\,\hat{d\mathbf{y}}_p
\biggr]^{-n/2},
\end{equation}
where $\hat{d\mathbf{y}}_p = \mathbf{B}_p ( {\mathbf{B}'_p}
\mathbf{B}_p )^{-1} {\mathbf{B}'_p}  d\mathbf{y}_p$. These formulae
extend to arbitrary variance functions $h$ by substituting
$\mathbf{B}\mapsto\mathbf{B} h^{1/2}$, $d\mathbf{y} \mapsto
d\mathbf{y} h^{1/2}$.
Network inference may now be carried out by Bayesian model averaging,
using the posterior probability of a directed edge from variable $i$
to variable $j$:
%
\begin{equation}
\mathbb{P}(i\mbox{ regulates }j)  =  \sum_G \frac{\pi(\mathbf{y}|G)
\pi(G)}{\sum_{G'} \pi(\mathbf{y}|G') \pi(G')} \mathbb{I}\{(i,j)\in
E(G)\}.
\end{equation}
In experiments below, we take a network prior which, for each variable
$p$, is uniform over the number of predictors $m_p$ up to a maximum
permissible in-degree $d_{\max}$, that is,
$\pi(G) \propto \prod_p {
P \choose m_p}^{-1} \mathbb{I}\{m_p \leq d_{\max}\}$, but note
that richer subjective network priors are available in the literature
[\citet{Mukherjee}].
Finally, a network estimator $\hat{G}$ is obtained by thresholding
posterior edge probabilities:
$(i,j)\in E(\hat{G}) \Leftrightarrow\mathbb{P}(i\mbox{ regulates }j)
> \varepsilon$.
For small maximum in-degree $d_{\max}$, exact inference by enumeration
of variable subsets may be possible. Otherwise, Markov chain Monte
Carlo (MCMC) methods can be used to explore an effectively smaller
model space [\citet{Ellis}; \citet{FriedmanIII}]. In the experiments
below we use exact inference by enumeration.

\subsubsection{Variable selection by corrected AIC}
Again, consider a constant variance function ($h= 1$); rescaling as
described above recovers the general case. The usual maximum
likelihood estimates
$\hat{A}_{p,\bullet} = ({\mathbf{B}'_p} \mathbf{B}_p)^{-1}
{\mathbf{B}'_p}  d\mathbf{y}_p$ and
$\hat{\sigma}_p^2 = \frac{1}{n} \sum_j (d\mathbf{y}_p(t_j)-
\hat{d\mathbf{y}}_p(t_j))^2$ induce closed forms $C_p\hat{\sigma}_p^{-
n}$ for the maximized factors of the likelihood function, where $C_p$
is a constant not depending on the choice of predictors.
Corrected AIC scores [\citet{Burnham}] for each variable $p$ are then
%
\begin{equation}
\operatorname{AIC}_c(p,G) = n\log(\hat{\sigma}_p^2) + 2m_p + \frac{2m_p(m_p+1)}{n-
m_p-1}.
\end{equation}
Again we consider all models with maximum permissible in-degree $d_{\max
}$. Lowest scoring models are chosen for each variable in turn,
inducing a network estimator~$\hat{G}$.

\section{Results} \label{sec: Results}

In this section we present empirical results investigating the
performance of a number of network inference schemes that are special
cases of the general formulation described by equation (\ref{eq: model}).
Objective assessment of network inference is challenging [\citet
{Prill}], since for
most biological applications the true data-generating network is
unknown.
We therefore exploit two published dynamical models of biological
processes, namely, \citet{Cantone} and \citet{Swat}, described in detail
in the Supplemental Information
[SI; \citet{Oates}].
The first is a synthetic gene regulatory network built in the yeast
\textit{Saccharomyces cerevisiae}. These five gene network
and associated delay differential equations (DDEs)
have received attention in computational biology
[\citet{Camacho}; \citet{Minty}], and have been shown to agree with gold-standard
data [at least under an
$\mathbb{E}(\mathbf{f}(\mathcal{F}_{\mathbf{X}})) \approx\mathbf
{f}(\mathcal{F}_{\mathbb{E}
(\mathbf{X})})$ assumption].
Cantone {et al.} consider two experimental conditions:
``switch-on'' and ``switch-off.'' In this paper ``switch-on'' parameter
values were used to generate data.
The Swat model is a gene-protein network governing the G$_1/$S
transition in mammalian cells. The model has a nine-dimensional state
vector and, unlike Cantone, is Markovian. We note that this model has
not been directly verified in the manner of Cantone but is based on a
theoretical understanding of cell cycle dynamics.
There is undoubtedly bias from this essentially arbitrary choice of
dynamical systems, but a comprehensive sampling of the (vast) space of
possible networks and dynamics is beyond the scope of this paper.

\begin{figure}
\centering
\begin{tabular}{@{}c@{}}

\includegraphics{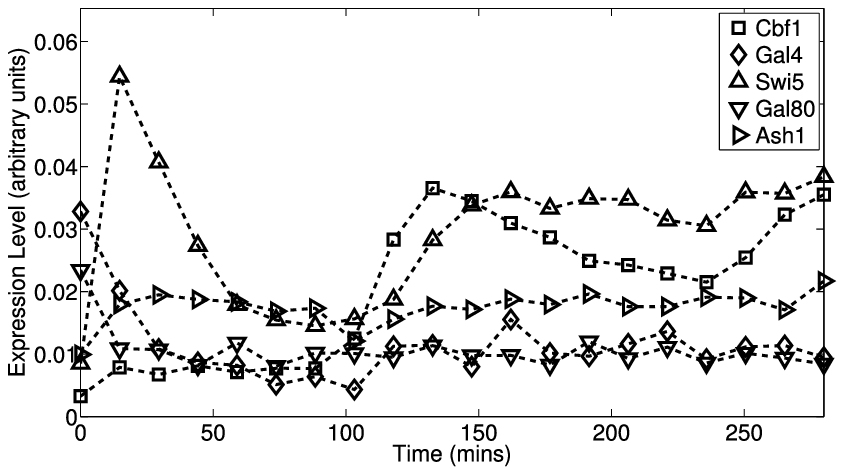}
\\
\footnotesize{(a)}\\[3pt]

\includegraphics{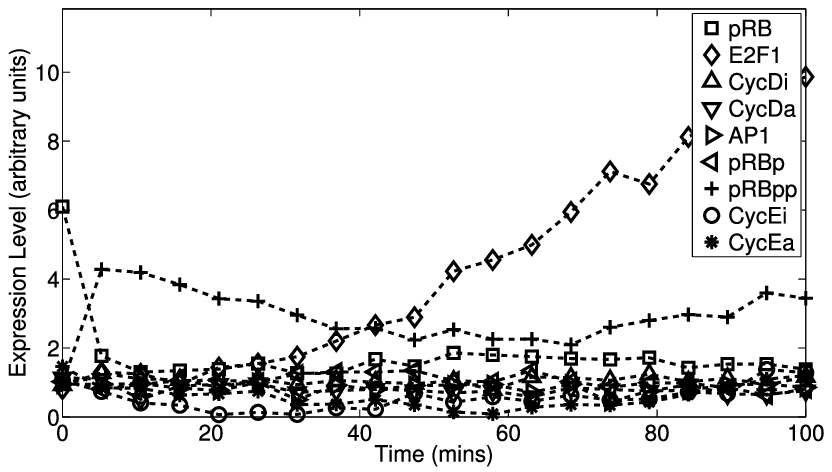}
\\
\footnotesize{(b)}
\end{tabular}
\caption{Two published dynamical systems models of cellular processes
were used to generate data sets. Single-cell trajectories were
generated from an SDDE model [equation (\protect\ref{eq: single cell})] and averaged
under measurement noise and nonlongitudinality due to destructive
sampling. (\textup{a}) Data generated from (a model due to) \protect\citet{Cantone},
describing a synthetic network built in yeast. (\textup{b}) Data
generated from \protect\citet{Swat}, a theory-driven model
of the G$_1/$S transition in mammalian cells.}
\label{fig: data}
\end{figure}

\subsection{Experimental procedure}

\subsubsection{Simulation}
We consider global perturbation data by initializing the dynamical
systems from out of equilibrium conditions. This is a common setting
for network inference approaches, but the limitations of inference from
such data remain incompletely understood.
For each dynamical system $\mathbf{f}$, trajectories $\mathbf{X}^k$ of
single-cell expression levels were obtained as solutions to the SDDE
equation (\ref{eq: single cell}) with drift $\mathbf{f}$ and uncorrelated
diffusion $\mathbf{g}(\mathbf{X}) = \sigma_{\mathrm{cell}} \mathcal{D}
(\mathbf{X})$ (representing multiplicative cellular noise).
Trajectories were obtained by numerically solving SDDEs with
heterogeneous initial conditions using the Euler-Maruyama
discretization scheme [equation (\ref{eq: Euler})]. MATLAB R2010a code for
all simulation experiments is available in the SI.
To mimic destructive sampling and consequent nonlongitudinality,
solutions were regenerated at each time point.
We are interested in data that are averages over a large number $N$ of
single-cell trajectories. However, the computational cost of solving
$N\times n$ SDDEs to produce each data set is prohibitive. Therefore,
only a smaller number $N^*\ll N$ of cells were simulated and a larger
sample $N$ then obtained by bootstrapping, that is, resampling from the
$N^*$ trajectories with replacement. In practice, $N^*$ should be taken
sufficiently large such that a negligible change in experimental outcome
results from further increase in $N^*$.
Initial conditions for single-cell trajectories varied with standard
deviation $\sigma_{\mathrm{cell}}$.
Finally, uncorrelated Gaussian noise of magnitude
$\sigma_{\mathrm{meas}}$ was added to simulate a measurement process
with additive error.
In the experiments presented below, $N = 10\mbox{,}000$, $N^*=30$ and
$n = 20$ time points are taken within the dynamically
interesting range (0--280 minutes for Cantone and 0--100 minutes for
Swat).
Measurement error and cellular noise are set to give signal-to-noise
ratios
$\langle\mathbf{X} \rangle/ \sigma_{\mathrm{meas}} \approx
10$, $\langle\mathbf{X} \rangle/ \sigma_{\mathrm{cell}}
\approx10$ [here $\langle\mathbf{X} \rangle$ represents
the average expression levels of the variables $\mathbf{X}$ over all
generated trajectories].
Figure~\ref{fig: data} shows typical data sets for the two dynamical
systems.

\subsubsection{Inference schemes}
The following inference schemes were assessed:\vspace*{6pt}

\begin{center}
\begin{tabular}{l|r}
Variable selection & \{ Bayesian, $\operatorname{AIC}_c$ \} \\
Design matrix & \{ Standard, Quadratic \} \\
Lagged predictors & \{ No, Yes \} \\
Variance function $h(\Delta) \propto\Delta^{-
\alpha}$ & $\alpha= $ \{ $0$, $1$, $2$ , $\varnothing$ \}\\
\end{tabular}
.
\end{center}\vspace*{6pt}

For the design matrix ``quadratic'' refers to the
augmentation of the predictor set by the pairwise products of
predictors, the simplest nonlinear basis functions. For the variance
function the symbol $\varnothing$ is used to denote the $\operatorname{VAR}(q)$ model,
which formally lacks a variance function.
``Lagged predictors${} = {}$Yes'' indicates augmentation of the predictor set
with lagged observations (a lag of $\approx28$ mins is used for
Cantone and $\approx10$ mins for Swat).
There are heuristic justifications for each of the candidate variance
functions. For example, the function with $\alpha= 2$ appears for
small $\Delta_j$ when an exact Euler approximation and additive
measurement error are assumed [\citet{Bansal}], whereas $\alpha= 1$ is
reminiscent of the Euler-Maruyama discretization equation (\ref{eq: Euler}).

\begin{figure}
\centering
\begin{tabular}{@{}c@{}}

\includegraphics{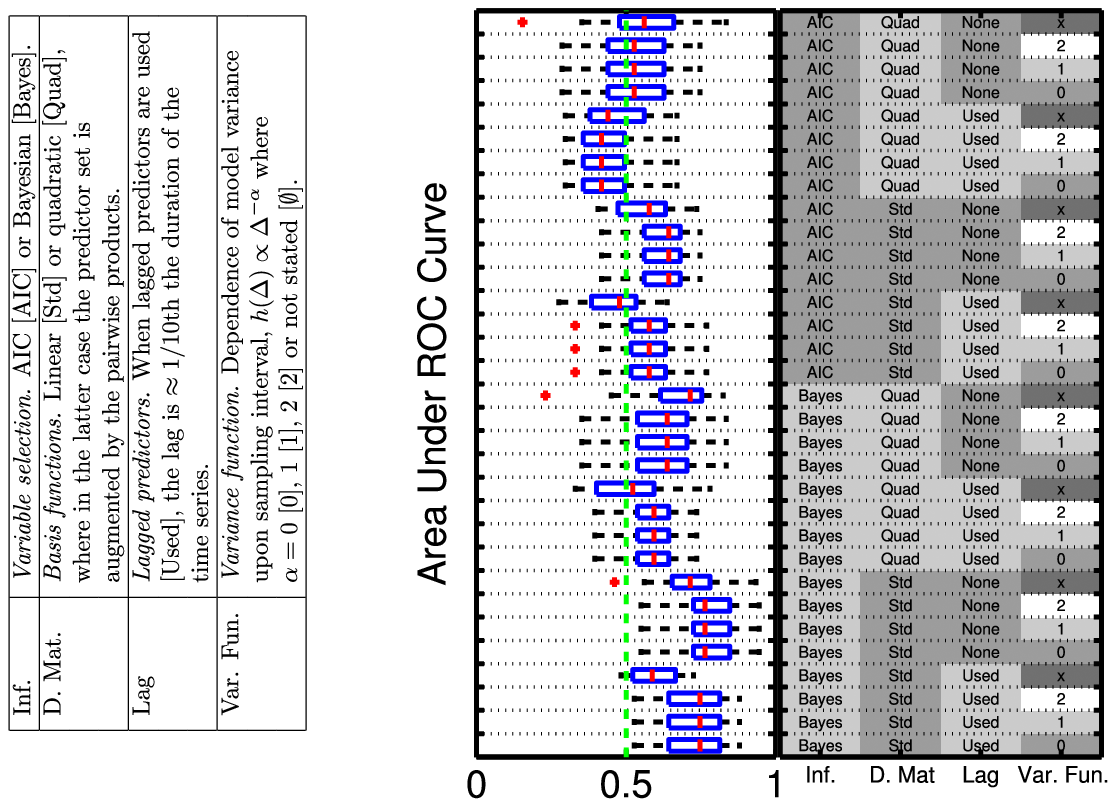}
\\
\footnotesize{(a)}\\[3pt]

\includegraphics{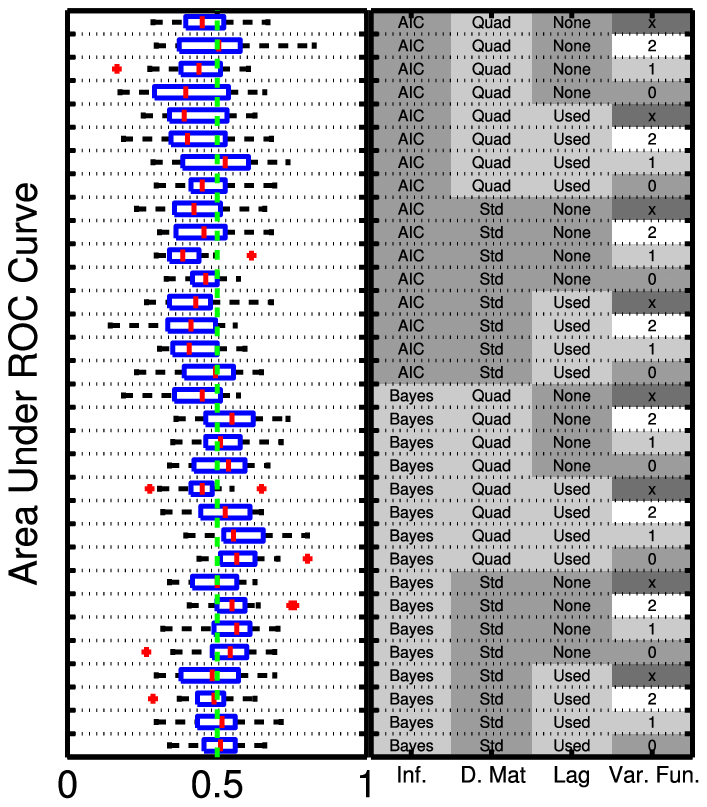}
\\
\footnotesize{(b)}
\end{tabular}
\caption{An empirical comparison of network inference schemes.
Simulated experiments based on published dynamical
systems allow benchmarking of performance in terms of area under ROC
curves (AUR; higher scores correspond to better network inference
performance). (\textup{a}) Even sampling intervals. (\textup{b})
Uneven sampling intervals.}
\label{fig: box}
\end{figure}

\begin{figure}[b]
\vspace*{-3pt}
\includegraphics{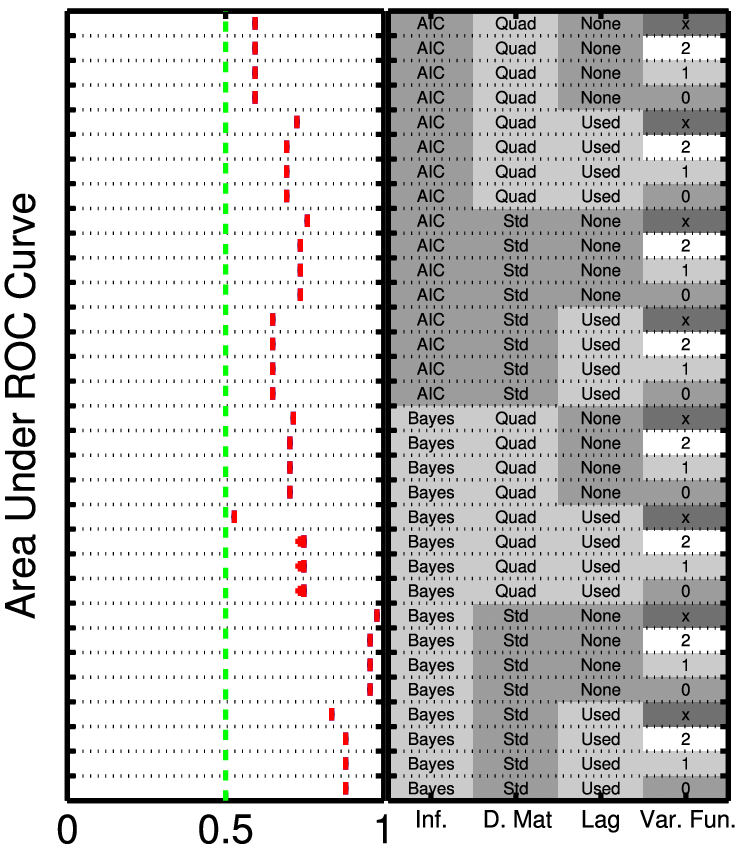}

\caption{Investigation of
empirical consistency of network estimators, using the \protect\citet{Cantone}
model with even sampling intervals. Area under ROC curves are
shown in the large data set, zero cellular heterogeneity and zero
measurement noise limits.}
\label{fig: consist}
\end{figure}

\subsubsection{Empirical assessment}

The performance of each inference scheme is quantified by the area
under the receiver operating characteristic (ROC) curve (AUR),
averaged over 20 data sets [\citet{Fawcett}]. This metric, equivalent to
the probability that a randomly chosen true edge is preferred by the
inference scheme to a randomly chosen false edge,
summarizes, across a range of thresholds, the ability to select edges
in the true data-generating graph.
Results presented below use a computationally favorable in-degree
restriction $d_{\max} = 2$. In order to check robustness to $d_{\max}$,
all experiments were repeated using $d_{\max}=3$, with no substantial
changes in observed outcome (SFigure 6).

\subsection{Empirical results}

\subsubsection{Even sampling interval}
Figure~\ref{fig: box}(a) displays box-plots over AUR scores for the
Cantone dynamical system under even sampling intervals.
Note that under even sampling, for an otherwise identical
scheme, changing variance function does not affect the model, leading
to identical AUR scores for schemes which differ only in variance
function. (An exception to this is the VAR model, since the parameters
$\mathbf{A}$ carry a subtly different meaning, which under a Bayesian
formulation leads to a translation of the prior distribution and in
the information criteria case changes the definition of the predictor
set.)

Despite the presence of nonlinearities and memory in the cellular
drift~$\mathbf{f}$, neither the use of quadratic basis functions nor the
inclusion of lagged predictors appear to improve performance in terms
of AUR. In order to verify that quadratic predictors are sufficiently
nonlinear and that lagged predictors are sufficiently delayed, we
repeated the investigation using both cubic predictors and using a
delay twice as long. Results (SFigures 3 and 4) demonstrate that no
improvement to the AUR scores is achieved in this way.

Corresponding results for the Swat model are shown in Figure~\ref{fig: box}. Here we
find that none of the methods perform well.

We also performed inference using biochemical data from the
experimental system reported in \citet{Cantone} (specifically the
``switch-on'' data set therein). AUR scores obtained using this data
(SFigure 5) were in close agreement with those obtained using synthetic
data [Figure~\ref{fig: box}(a)], suggesting that the results of the
simulations are relevant to real world studies.

\subsubsection{Uneven sampling intervals}
Many biological time-course experiments are carried out with uneven
sampling intervals.
We therefore repeated the analysis above with sampling times of 0, 1,
5, 10, 15, 20, 30, 40, 50, 60, 75, 90, 105, 120, 140, 160, 180, 210,
240 and 280 minutes.
Figure~\ref{fig: box}(b) displays the AUR scores so obtained.
We find that all the methods perform worse in the uneven sampling
regime, with no method performing significantly better than random.
Corresponding results for the Swat model are shown in SFigure 7. Again,
here we find that none of the methods perform well.\looseness=-1

\subsubsection{Consistency}
Figure~\ref{fig: consist} displays AUR scores for Cantone for a large
number of evenly sampled time points ($n = 100$), and the limiting
case of zero measurement noise and zero cellular heterogeneity
($\sigma_{\mathrm{meas}} = 0$, $\sigma_{\mathrm{cell}} = 0$, even sampling
intervals).
Consistency
(in the sense of asymptotic convergence of the network estimate
to the data-generating network)
may be unattainable due to the nonidentifiability resulting from
limited exploration of the dynamical phase space. This lack of subjectivity
means that in many cases inference cannot possibly reveal
the full data-generating graph, although, as we have seen,
network inference can nonetheless be informative.
From Figure~\ref{fig: consist} we see that the Bayesian schemes using
linear predictors approach AUR equal to unity, and in this
sense show empirical consistency with respect to network inference.
However, some of the other methods do not converge to the correct
graph even in this limit.

\section{Discussion} \label{sec: discuss}

The analyses presented here were aimed at
better understanding statistical network inference
for biological applications. We showed how a broad class of
approaches, including VAR models, linear DBNs and certain
ODE-based approaches, are related to stochastic dynamics at the cellular
level.
We discuss a number of these aspects below and close with some views on
future perspectives for network inference,
including recommendations for practitioners.

\subsection{Time intervals} We found that uneven sampling intervals
posed problems, even for
methods that explicitly accounted for the sampling interval.
Further insight may be gained from a ``propagation of uncertainty''
analysis of the
approximations indicated in Section~\ref{sec: assumptions}. Assuming
the true large sample process obeys $d\mathbf{X}^\infty/dt =
\mathbf{F}(\mathbf{X}^\infty)$, we have that under an observation
process with independent additive Gaussian
measurement error $\mathbf{Y}(t) \sim N(\mathbf{X}^{\infty}(t),\mathbf
{M})$ an expansion for
the variance $\mathbb{V}(d\mathbf{Y}-\mathbf{F}(\mathbf{Y}
))$ over a~time interval $\Delta$ is given by
%
\begin{equation}
\mathbf{M}\Delta^{-2} + (\mathbf{I}\Delta^{-1}+D\mathbf{F})
\mathbf{M} (\mathbf{I}\Delta^{-1}+D\mathbf{F})' + \cdots\label{eq:
htrue}
\end{equation}
(see SI for details). Recall that the model family in equation (\ref{eq: model})
approximates this variance by $h(\Delta)\mathcal{D}(\sigma_1^2,
\ldots, \sigma_P^2)$, where $h(\Delta) = \Delta^{-\alpha}$. From this
perspective it is clear that each
variance function we considered captures only partial variation due to
$\Delta$. It is therefore not surprising that
performance suffers in the uneven sampling regime, which requires the
variance function to apply equally to large $\Delta$ as to small
$\Delta$. Moreover, a natural choice of variance function driven by
equation (\ref{eq: htrue}) is not possible, since this would require
knowledge of the unknown process $\mathbf{F}$. The implication for
experimental design is that, absent specific reasons for uneven
sampling, it may be preferable to collect data at regular intervals.

\begin{figure}

\includegraphics{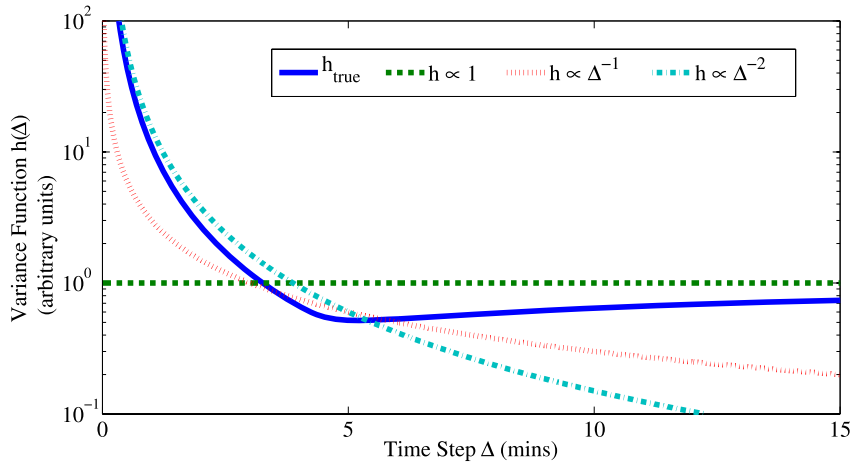}

\caption{Variance functions used in literature provide partial
approximation to the ``true'' functional form for \protect\citet{Cantone}. For
small time steps a power law $\Delta^{-\alpha}$
provides a good approximation, but for larger time steps a constant
variance function may be more appropriate. In practice, the precise
form of
$h_{\mathrm{true}}$ will be unknown.}
\label{fig: variance functions}
\end{figure}

Figure~\ref{fig: variance functions} displays an approximation to the
true variance function for the Cantone model (see SI). Observe that
for small sampling intervals $\Delta$ the true curvature is best
captured by a functional approximation of the form $h(\Delta) \propto
\Delta^{-\alpha}$ with $\alpha= 1,2$, whereas for intervals larger
than 10 mins (which are more common in practice) the flat
approximation $h(\Delta) \propto1$ correctly captures the asymptotic
behavior. In applications where high frequency sampling is infeasible,
the flat variance function might be a
sensible choice.
To understand whether difficulties related to sampling intervals
disappear in the large sample\vadjust{\goodbreak} limit, we repeated the empirical
consistency analysis under uneven sampling (SFigures 11 and 12).
Interestingly, we found that none of the methods appeared to be
empirically consistent, and that the choice of variance function is influential.
However, unevenly sampled data are common in biology and it may be the
case that in some settings, the existence of multiple time scales
(e.g., signaling, transcription, accumulating epigenetic alterations)
mean that unevenly sampled data are nonetheless useful.
Our findings suggest that care should be taken in the
uneven sampling regime.

\subsection{Interventional data}
The Cantone data are favorable in the sense that gene profiles show
interesting time-varying behavior under global perturbation, exploring
a large proportion of the dynamical phase space.
However, such behavior is dependent on the specific dynamical system
and is not displayed by the Swat model, which has a much larger phase
space, being a~nine-dimensional dynamical system.
This may help explain the poor performance of all the methods on this
latter model using global perturbation data and
perhaps reinforces the intuitive notion that
dynamics that are favorable (in this informal sense) facilitate
network inference.
In some cases, perturbation data are available in which individual
variables are inhibited (e.g., by RNA interference, gene knockouts or inhibitor
treatments). Such data have the potential to explore much more of the
dynamical phase space, including regions which cannot be accessed
without direct inhibition of specific molecular components. This is an
important consideration because the statistical estimators described in
Section~\ref{sec: inference} take the form
%
\begin{equation}
\hat{\mathbf{A}} = \langle D\mathbf{f}(\mathcal{F}_{\mathbf{X}}) \rangle
_{\mathbf{X} \in\mathcal{R}},
\label{eq: ave est}\vadjust{\goodbreak}
\end{equation}
where the average is over the region $\mathcal{R}\subseteq\mathcal{X}$
in state space visited during the experiments. Clearly, if the region
$(\mathbf{f}(\mathcal{F}_{\mathcal{R}}),\mathcal{R})$
is only a small subspace of phase space, then the estimate equation (\ref
{eq: ave est}) will be poor compared to one based on the entire phase
space $\hat{\mathbf{A}}^* = \langle D\mathbf{f}(\mathcal{F}_{\mathbf
{X}}) \rangle_{\mathbf{X} \in\mathcal{X}}$.

To investigate the added value of interventional treatments for network
inference,
we repeated both the Cantone and Swat analyses with an ensemble
of data sets obtained by inhibiting each variable in turn;
this gave 5 and 9 data sets for Cantone and Swat respectively.
While no improvement to the Cantone AUR scores was observed (SFigure
15), there was improved performance for Swat (SFigure 16). This
suggests that global perturbations are insufficient to explore the Swat
dynamical phase space, and supports the intuitive notion that
intervention experiments may be essential for inference regarding
larger dynamical systems. Nevertheless, AUR scores remain far from
unity. This may be because the Swat drift function contains complex
interaction terms which single interventions alone fail to elucidate.
An important problem in experimental design will be to estimate how
much (possibly combinatorial) intervention is required to achieve a
certain level of network inference performance.

We considered precise artificial intervention of single components
{in silico}. However, biological interventions may be imprecise and
imperfect. For example, RNA interference achieves only incomplete
silencing of the target and small molecule inhibitors may have
off-target effects. Moreover, at present such interventions are not
instantaneous nor truly exogenous. This means that in many cases the
system itself may be changed by the intervention, rendering resulting
predictions inaccurate for the native system of interest.
There remains a need for novel statistical methodology capable of
analyzing time-course data under biological interventions.
Existing literature in causal inference [\citet{Pearl}] and related work
in graphical models [\citet{Eaton}] are relevant, but in biological
applications it may also be important to consider the mechanism of
action of specific interventions.

\subsection{Nonlinear models}
We focused on linear statistical models.
Clearly, linear models are inadequate in many cases.
For example, \citet{Rogers} demonstrate the benefit of a nonlinear model
based on Michaelis-Menten chemical kinetics
for inference of transcription factor activity.
However, network inference based on nonlinear ODEs remains challenging
[\citet{Xu}].
Alternatively, \citet{Aijo} consider the use of a nonparametric Gaussian
process (GP) interaction term in the regression, which is naturally
more flexible than linear regression using finitely many basis
functions. This may help to overcome the linearity restriction, but
introduces additional degrees of freedom, including the GP covariance
function and associated hyperparameters. While a thorough comparison of
such approaches was beyond the scope of this article, the potential
utility of nonparametric interaction terms is worthy of investigation.
In this study we observed that neither the use of predictor products
nor lagged predictors led to improved performance; this may reflect
nontrivial coupling between cellular dynamics and the observed data.

\subsection{Single-cell data}
In the future it may become possible to measure single-cell expression
levels $\mathbf{X}^k$ nondestructively (e.g., by live cell imaging),
producing truly longitudinal data sets.
It is interesting to consider how such data may impact upon the
performance of regression-based network inference. Under independent
additive Gaussian measurement error $\mathbf{Y}(t) \sim N(\mathbf
{X}^k(t),\mathbf{M})$ an expansion for the single-cell variance $\mathbb
{V}(d\mathbf{Y}-\mathbf{f})$ over a time interval $\Delta$, in analogy
with equation~(\ref{eq: htrue}), is given by
%
\begin{equation}
\mathbf{M}\Delta^{-2} + (\mathbf{I}\Delta^{-1}+D\mathbf{F})
\mathbf{M} (\mathbf{I}\Delta^{-1}+D\mathbf{F})' + \Delta^{-1}\mathbf
{gg}' + \cdots
\end{equation}
(see SI). Thus, a (single) longitudinal single-cell data set contains
less information about the drift $\mathbf{f}$ than aggregate data
[equation (\ref{eq: htrue})] due to cellular stochasticity~$\mathbf{g}$.
However, multiple longitudinal data sets may jointly contain more
information than a single aggregate data set.
To empirically test the utility of such data, we carried out network inference
using 10 such longitudinal single-cell data sets on both the Cantone
and Swat models, observed at even intervals with the same magnitude of
measurement error as aggregate data.
Results (SFigures 13 and14) show a small improvement to the mean AUR
scores, but reduction by a factor of about two in the variance of
these scores (compared with the corresponding nonlongitudinal data),
implying that the network estimators may be converging to an incorrect
network. Bias may occur when the cellular drift $\mathbf{f}$ is not
well approximated by a linear function, as is the case for the Swat
model. Consider the idealized scenario where $\mathbf{f}\equiv\mathbf
{f}(\mathbf{X})$ is Markovian and it is possible to observe
longitudinal, single-cell expression levels. Under these apparently
favorable circumstances even estimators obtained after a thorough
exploration of state space may not offer good approximations, that is,
$\hat{\mathbf{A}}^* \not\approx D\mathbf{f} |_{\mathbf
{x}=\mathbf{0}}$. As a toy example consider the cellular drift
%
\begin{equation}
\mathbf{f}:[0,1]^2 \rightarrow\mathbb{R},   \qquad  \mathbf{f}(\mathbf
{X}) = \left(
\matrix{(2\pi)^{-1}\operatorname{sin}(2\pi X_2) \vspace*{2pt}\cr
(2\pi)^{-1}\operatorname{sin}(2\pi X_1)}
\right),
\end{equation}
which is not well approximated by a linear function over the state
space $\mathcal{X} = [0,1]^2$. In this case averaging leads to cancellation
%
\begin{eqnarray}
\hat{\mathbf{A}}^* &=& \langle D\mathbf{f}(\mathbf{X}) \rangle_{\mathbf
{X}\in\mathcal{X}}  =  \left\langle
\pmatrix{ 0 & \operatorname{cos}(2\pi X_2) \vspace*{2pt}\cr \operatorname{cos}(2\pi X_1) &
0}
\right \rangle_{\mathbf{X}\in[0,1]^2}
\nonumber
\\[-8pt]
\\[-8pt]
\nonumber
& = & \mathbf{0} \neq
\pmatrix{ 0 & 1 \vspace*{2pt}\cr 1 & 0}= D\mathbf{f}|_{\mathbf{x}=\mathbf{0}}
\end{eqnarray}
so that no interactions are inferred. Under such circumstances network
inference is no longer possible using the na\"{i}ve linear regression approach.
This suggests that network inference rooted in nonlinear models
may be needed to fully exploit
longitudinal single-cell data in the future.
A related line of work addresses heterogeneity
of the drift function in time by coupling DBNs with change point
processes [\citet{Grzegorczyk}; \citet{Kolar}; \citet{Lebre}]. A promising direction would
be piecewise linear regression modeling for network inference
applications, where the heterogeneity appears in the spatial
domain.\vspace*{-3pt}

\subsection{High-dimensions and missing variables}
We focused on the simplest possible case of fully observed,
low-dimensional systems.
There is a rich literature in high-dimensional variable selection and
related graphical models [\citet{Meinshausen}; \citet{Hans}; \citet{FriedmanII}] which
applies equally to the regression models described here.
The issues raised in this paper
remain relevant in the high-dimensional setting.
However, in practice, even high-dimensional observations are likely to
be incomplete, since
it is not currently possible to measure all relevant chemical species.
Therefore, inferred relationships between variables may be
indirect. This may be acceptable for the purpose of predicting the
outcome of biochemical interventions (e.g., inhibiting gene or protein
nodes), but limits stronger
causal or mechanistic interpretations.
Latent variable approaches are available [\citet{Beal}], but model
selection can be challenging and remains an open area of research
[\citet{Knowles}].
We note also that the missing variable issue for biological networks
is arguably more severe than in, say, economics or epidemiology,
insofar as measured variables may represent only a small fraction of
the true state vector, often with little specific insight available
into the nature of the missing variables or their relationship to
observations.
Further work is required to better understand these issues in the
context of inference for biological networks.\vspace*{-3pt}

\subsection{Future perspectives}

We found that a simple linear model could successfully infer network
structure using globally perturbed time-course data
from the Cantone system. It is encouraging that inference based only on
associations between variables, none of which were explicitly
intervened upon, can in some cases be effective.
Interventional designs should further enhance prospects for network
inference. On the other hand, theoretical arguments, and the results we
showed from the Swat system, emphasize that in some cases network
structure may not be identifiable, even at the coarse level required
for qualitative biological prediction.
On balance, we believe that network inference can be useful in
generating biological hypotheses and guiding further experiment.
However, the concerns we raise motivate a need for caution in
statistical analysis and interpretation of results. At the present
time, we\vadjust{\goodbreak} do not believe network inference should be treated as a
routine analysis in bioinformatics applications, but rather as an open
research area that may, in the future, yield standard experimental and
statistical protocols.

Some specific recommendations that arise from the results presented
here are as follows:
\begin{itemize}
\item\textit{A default model.} Our results suggest that a reasonable
default choice
of model for typical applications uses
the standard design matrix with no lagged predictors and a flat
variance function, corresponding to the linear model
%
\begin{equation}
d\mathbf{Y}(t_j) \sim N(\mathbf{A} \mathbf{Y}(t_{j-1}),\mathcal{D}
(\sigma_1^2,\ldots,\sigma_P^2)).
\end{equation}

Coupled with the Bayesian variable
selection scheme outlined in Section~\ref{subsubsec: Bayes}, this simple
model produced empirically consistent network estimators for Cantone
using evenly sampled global perturbation data (Figure~\ref{fig: consist}).

\item\textit{Diagnostics and validation}.
It is clear that network inference does not enjoy general theoretical
guarantees and that the ability to successfully elucidate network
structure depends on details of the specific system under study.
Therefore, careful empirical validation on a case-by-case basis is
essential. This should include statistical assessment of model fit,
robustness and predictive ability and, where possible, systematic
validation using independent interventional data.

\item\textit{Experimental design}. We suggest sampling evenly in time as
a default choice. Interventional designs may be helpful to effectively
explore larger dynamical phase spaces. However, to control the burden
of experimentally exploring multiple time points, molecular species,
interventions, culture conditions and biological samples, adaptive
designs that prune experiments based on informativeness for the
specific biological setting may be helpful [\citet{Xu}].
\end{itemize}

In conclusion, linear statistical models for networks are closely
related to models of cellular dynamics and can shed light
on patterns of biochemical regulation.
However, biological network inference remains profoundly
challenging, and in some cases may not be possible even in principle.
Nevertheless, studies
aimed at elucidating networks from high-throughput data are now
commonplace and play a prominent role in biology.
For this reason there remains an urgent need for both new methodology
and theoretical and empirical investigation of existing approaches.
Furthermore, there remain many open questions in experimental design
and analysis of designed experiments in
this setting.

\section*{Acknowledgments}
We would like to thank Professor K. Kafadar and the anonymous referees
for constructive suggestions that helped to improve the content and
presentation of this article, and G. O. Roberts, S. Spencer and
S. M. Hill for discussion and comments.\vadjust{\goodbreak}

\begin{supplement}[id=suppA]
\stitle{Additional materials}
\slink[doi]{10.1214/11-AOAS532SUPP} 
\slink[url]{http://lib.stat.cmu.edu/aoas/532/}
\sdatatype{.zip}
\sdescription{This supplement provides the dynamical systems used in
this paper and accompanying MATLAB R2010a scripts, derivations and
additional figures SFigures~1--16.}
\end{supplement}



\printaddresses

\end{document}